\documentclass[prl,superscriptaddress,twocolumn,showpacs]{revtex4-1}
\usepackage{epsfig,amsmath,amssymb,graphics,color,calc}
\newcommand{\be}{\begin{equation}}
\newcommand{\ee}{\end{equation}}
\newcommand{\ba}{\begin{eqnarray}}
\newcommand{\ea}{\end{eqnarray}}

\begin{document}

\title{Ultra-long-range dynamic 
correlations in a microscopic model for aging gels}

\author{Pinaki Chaudhuri}

\affiliation{The Institute of Mathematical Sciences, C.I.T. Campus, 
Taramani, Chennai 600 113, India}

\author{Ludovic Berthier}

\affiliation{Laboratoire Charles Coulomb, UMR~5221, 
Universit\'e Montpellier and CNRS, 34095 Montpellier, France}

\date{\today}

\begin{abstract}
We use large-scale computer simulations to explore the nonequilibrium
aging dynamics in a microscopic model for colloidal gels. We find
that gelation resulting from a kinetically-arrested phase separation
is accompanied by `anomalous' particle dynamics revealed by
superdiffusive particle motion and compressed exponential relaxation
of time correlation functions. Spatio-temporal analysis of the
dynamics reveals intermittent heterogeneities producing spatial
correlations over extremely large length scales. Our study is the
first microscopically-resolved model reproducing all features of
the spontaneous aging dynamics observed experimentally in 
soft materials.
\end{abstract}

\maketitle

Understanding the complex mechanisms underlying the formation and
stability of colloidal gels remains a challenge, despite the diversity
of existing applications exploiting their mechanical properties.
Gels are low-density structures forming percolating networks, with
bonds that are either permanent (chemical gels) or transient (physical
gels) \cite{zaccarellirev,tombook,larsonbook,delgado,pablo, pre2010}.
A prevalent method to form physical gels follows a nonequilibrium
route by quenching a homogeneous fluid into a phase coexistence
region \cite{zaccarellirev, dave, cardinaux,lunature2008, paddynmat2008,
koyamaprl2009, paddy2}, which generates bicontinuous structures.
When the dense phase forms an amorphous solid, the phase separation
is kinetically hindered \cite{foffi1,foffi2,testard1,testard2} and
a gel forms.  The microscopic dynamical and structural properties
of these nonequilibrium gels evolve slowly with time.

This aging dynamics has been the subject of many experimental
studies, which established that aging in colloidal gels is `anomalous'
\cite{reviewluca,luca-ramos}, i.e.  it differs qualitatively from
the aging observed in conventional glassy materials, such as polymer
and colloidal glasses \cite{weeks-pre-2008,yodh-prl-2009}.  Scattering
experiments \cite{lucaprl2000,duriepl2006, ranjiniprl2004,bob2,madsen}
report that time correlation functions are described by {\it
compressed exponential} relaxations. Such behaviour differs from
the (exponential) diffusive dynamics in simple liquids or the
(stretched exponential) decay observed in glassy fluids~\cite{rmp}.
In addition, compressed exponentials are seen to emerge only for
{\it large enough displacements}, with the relaxation timescale
$\tau(q)$ crossing over from $\tau \sim{q^{-2}}$, characteristics
of diffusion, to $\tau \sim q^{-1}$, characteristic of ballistic
dynamics, with decreasing the scattering wavevector $q$. Finally,
spatially-resolved dynamic measurements revealed the existence of
{\it long-ranged correlations} extending up to the system size
\cite{duriprl2009,macsm2010}, again contrasting with the much
smaller-ranged dynamic heterogeneity observed in glassy materials
\cite{book}.  Such peculiar behaviour is hypothesized to follow
from the infrequent release of `internal stresses' that relax the
fractal network \cite{luca-ramos}, but this interpretation remains
to be confirmed by direct observation. This overall phenomenology
has been reported for laponite, carbon black, micellar polycrystals,
multilamellar vesicles, implying it is  generic to a large class
of soft materials \cite{luca-ramos}.

A microscopic perspective via theoretical modeling is also missing.
To study this problem, one needs a model with a realistic gel
structure, studied for large enough timescales to observe the slow
dynamics, and over large enough lengthscales to detect long-ranged
correlations. Our work successfully addresses these three challenges.
There have been extensive studies of aging effects in particle-based
computer simulations of dense glassy systems (e.g.,
\cite{kobbarrat,elmasri}), but none of the above anomalous
signatures are observed. For nonequilibrium gelation, the focus has
been mainly on the formation of arrested fractal structures via
quenches into the phase-coexistence region, with little insight
into the aging dynamics.  Compressed exponentials in time correlation
functions were reported in equilibrium gels but they result from
inertial effects at short time and length scales \cite{saw1,saw2}.
Other attempts produced compressed relaxations in gel-like structures
either by heating out-of-equilibrium initial states \cite{suarez}
or by driving the material mechanically \cite{ema}. 
Compressed relaxations were also reported for
mechanically induced coarsening at zero-temperature  
\cite{tanaka_epl}, but the lengthscale dependence and spatial
correlations were not analysed. A more recent work on a low density
gel reported instead subdiffusive behaviour \cite{zia}. At a more
coarse-grained level, it was suggested that localised bond-breaking
events in gels may result in compressed exponentials \cite{pitard},
a mean-field scenario that was recently revisited using a mesoscopic
elasto-plastic model for generic glassy materials \cite{ferrero2014},
in which the gel structure however plays no direct role.

Using large-scale numerical simulations of a particle-based model
for gel formation \cite{testard1,testard2}, we show that the
spontaneous microscopic aging dynamics during gelation possesses
all anomalous signatures reported experimentally.  We find a
subdiffusive aging dynamics at short lengthscales, corresponding
to caged particle motion inside the gel strands, crossing over to
superdiffusive relaxation at larger lengthscales triggered by
intermittent snapping of the fractal network.  These relaxation
events result in compressed exponential relaxations at small enough
scattering vectors and correspond also to dynamic correlation
lengthscales that are much larger than the typical pore size of the
gel. These results thus provide a coherent microscopic picture of
the anomalous aging dynamics observed experimentally.

Following previous work \cite{testard1,testard2}, we study the
properties of a binary Lennard-Jones mixture after it is suddenly
quenched into its liquid-gas phase coexistence region. To access
large-enough system sizes, we work in two spatial dimensions.  We
consider a 65:35 A-B Lennard-Jones mixture with model parameters
as in Ref.~\cite{bruning2009}.  Because of its glass-forming ability,
phase separation is kinetically hindered, which results in gelation.
We perform molecular dynamics simulations using LAMMPS \cite{plimpton95}.
In our simulations, energies and lengths are expressed in Lennard-Jones
units of interaction energy $\epsilon_{\textrm{AA}}$ and diameter
$\sigma_{\textrm{AA}}$ for the majority specie.  Particles have
equal masses, $m$, and the time unit is
$\sqrt{{m\sigma_{\textrm{AA}}^{2}}/\epsilon_{\textrm{AA}}}$.

In these reduced units, we observe incomplete phase separation of the binary
mixture when temperature is lower than $T \approx 0.3$, and a
bicontinuous fractal structure is obtained for number densities in
the range $0.25 < \rho < 1.1$.  Our strategy is to first equilibrate
the system at high temperature in the homogeneous region, $T=3.0$,
for different densities $\rho=0.4$, 0.5, 0.6, 0.7, 0.8, 1.0. Then,
at time $t=0$, we instantaneously quench the system to low temperatures
in the phase coexistence region, $T=0.05$, 0.10, 0.15, 0.20, 0.30,
0.40, and 0.50. Thereafter, we follow the dynamical properties of the
system at these different state points.  In order to explore spatial
correlations we study large systems ranging from $N=10^4$ to
$2\times{10^6}$ particles. A majority of the results shown below
correspond to $N=5{\times}10^5$.  All data are averaged over at
least 4 different initial conditions.  During the dynamics, the
temperature control is done via dissipative particle dynamics
(DPD) thermostat \cite{sk03}, using a friction coefficient value
of $\xi=1$; tests performed with larger damping $\xi=3$, $10$ 
showed identical results.

\begin{figure}
\includegraphics[width=8.5cm]{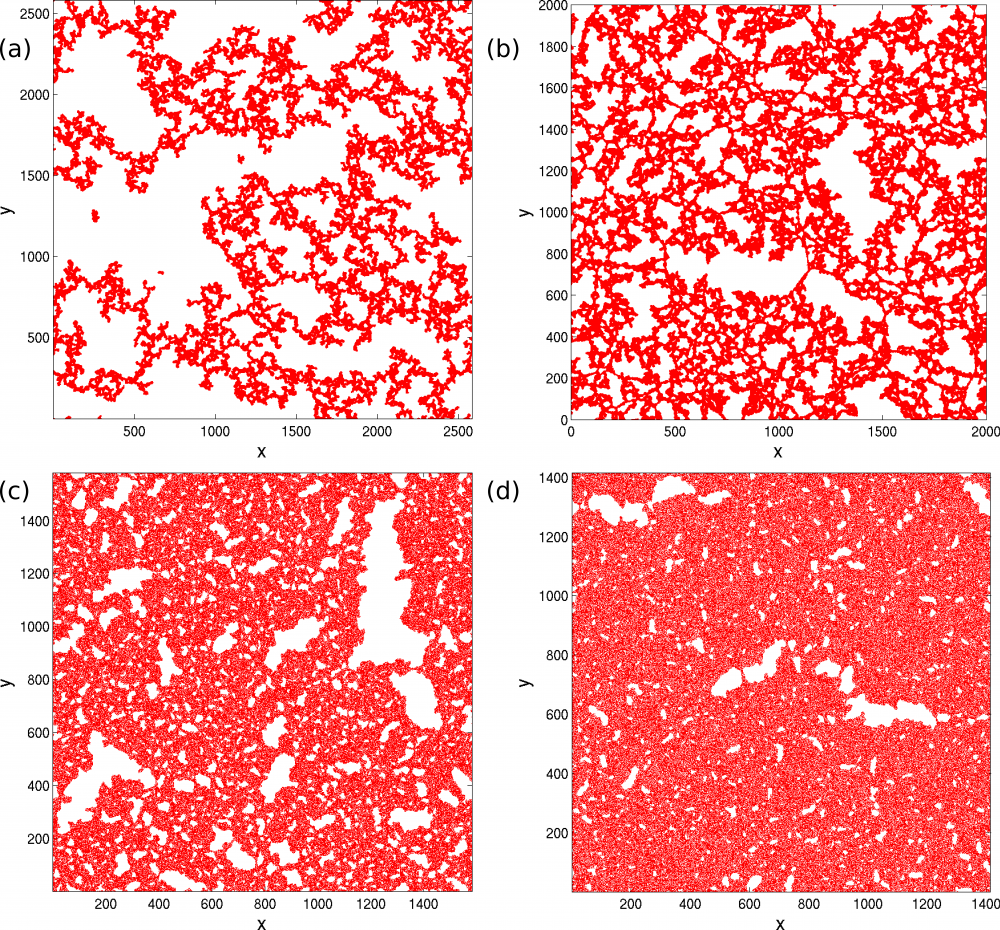}
\caption{Snapshots of the two-dimensional binary Lennard-Jones 
mixture at large time, $t=2.7\times{10^4}$, following
a quench from $T=3.0$ to $T=0.15$ at various 
densities 
(a) $\rho =0.3$, 
(b) $\rho =0.5$, 
(c) $\rho =0.8$, 
(d) $\rho =1.0$.}
\label{fig1}
\end{figure}

The nearly-arrested bicontinuous structures formed at large times
for $T=0.15$ and various densities are shown in
Fig.~\ref{fig1}, for a system with $N=2 \times {10^6}$. These images
reveal the coexistence of a connected network of dense amorphous
domains and void-like regions. For $\rho=0.3$, 0.5, the denser
regions form a fractal network with tenuous links, having a
morphology reminiscent of those formed during nonequilibrium
gelation in experiments. At higher densities, $\rho=0.8$,
1.0, the voids shrink and the dense domains become more
compact. At higher temperatures, $T > 0.3$, these structures would
rapidly coarsen until phase separation is complete.  On the other
hand at very low temperatures, the gelation process is essentially
arrested, because the particle bonds become nearly permanent.  A
detailed characterization of the gel structure can be found in
\cite{testard2}.  Below we explore the intermediate, experimentally
relevant regime where the gel evolves very slowly with time,
focusing on the temperature $T=0.15$.

\begin{figure}
\includegraphics[width=8cm]{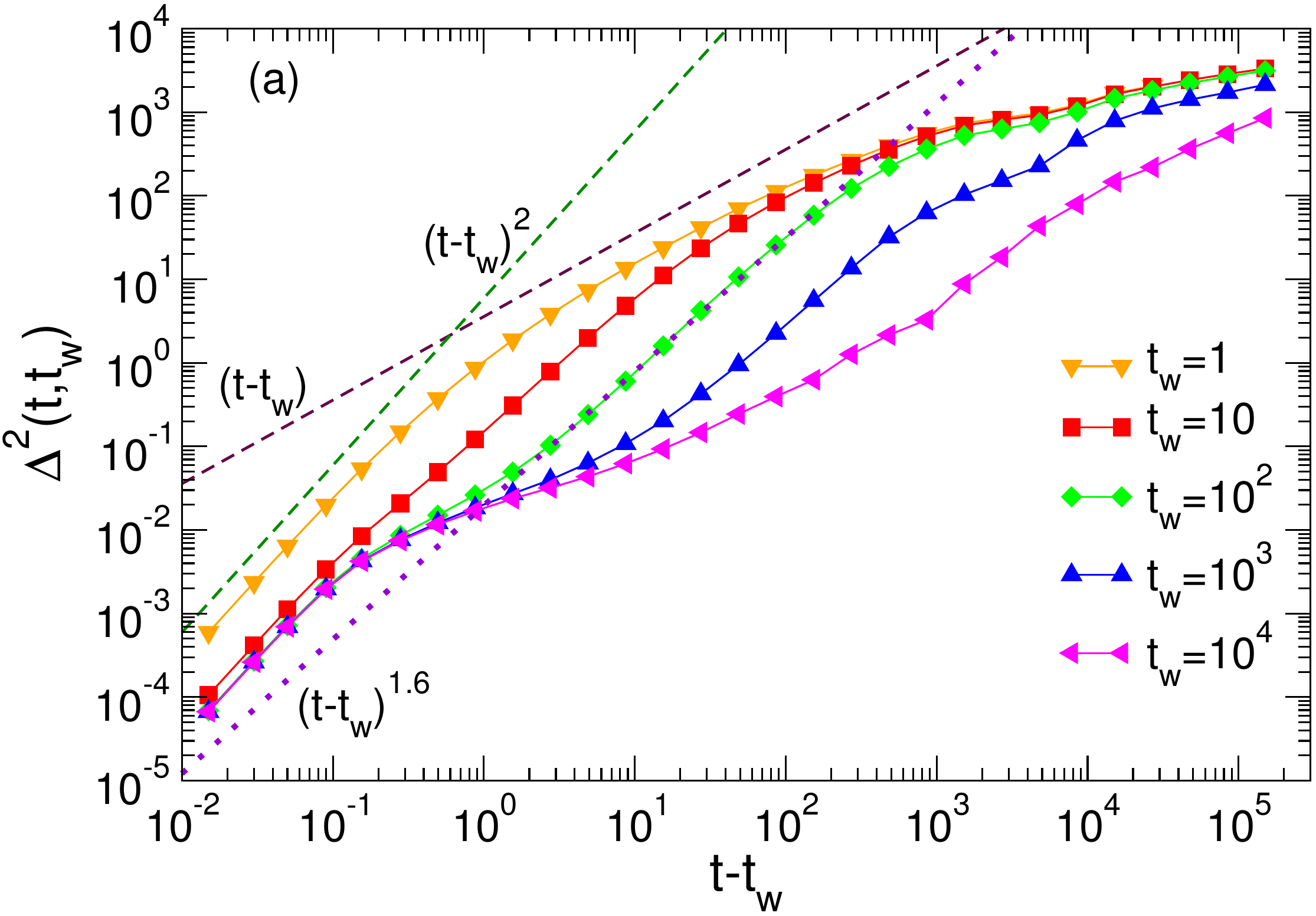}
\includegraphics[width=8cm]{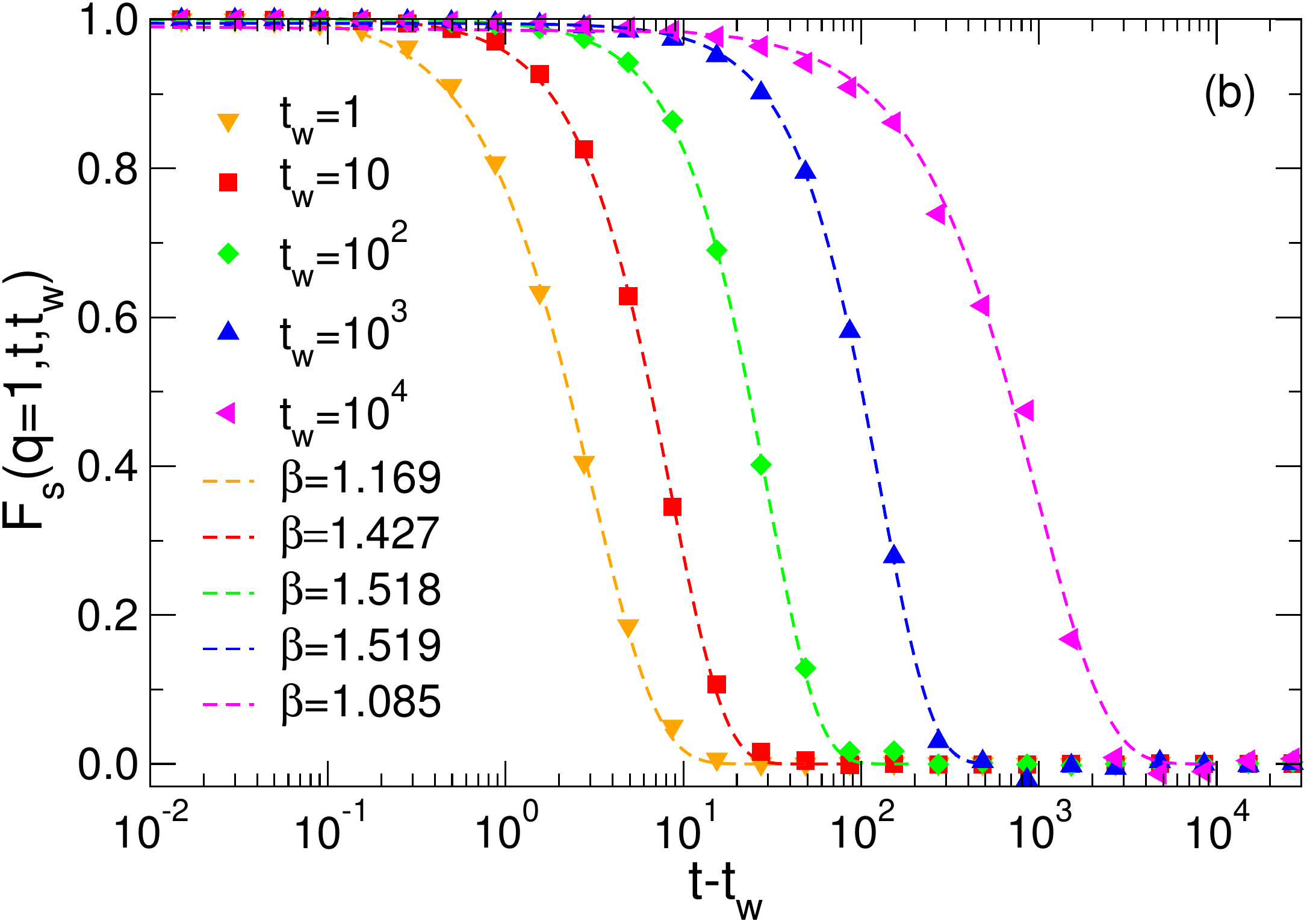}
\caption{$\rho=0.5$, $T=0.15$.
(a) Mean-squared displacements, $\Delta^2(t,t_w)$, 
measured for different ages, $t_w$.
A succession of ballistic, subdiffusive caging,
superdiffusive cage escape and long-time diffusion 
are observed, with various time dependences indicated via dashed lines. 
(b) Corresponding self-intermediate scattering function, $F_s(q,t,t_w)$, 
measured for $q=1$. 
The dashed lines correspond to fits with compressed exponentials 
with the corresponding $\beta$ values indicated.}
\label{fig2}
\end{figure}

To account for experimental findings, we need to study the microscopic
dynamics as a function of the time $t_w$ spent since gelation 
started. Such $t_w$-dependence 
is demonstrated in Fig.~\ref{fig2}(a) for
the mean-squared displacements, $\Delta^2(t,t_w) = \frac{1}{N_A}
\left\langle |{\bf r}_i (t) - {\bf r}_i(t_w)  |^2 \right\rangle$,
for the majority specie (type-A particles), where ${\bf r}_i(t)$
represents the position of particle $i$ at time $t$, and the brackets
represent an average over initial conditions.  The data in
Fig.~\ref{fig2}(a) reveal that particle motion depends
on $t_w$, the dynamics being slower for `older' systems,
a feature commonly observed in glassy materials.  The details of
the particle dynamics are however intriguing.  Except for the very
short waiting times where the gel structure is barely formed, four
dynamic regimes are observed.  For $t_w=10^2$ and $10^3$, for
instance, the initial ballistic regime is followed by subdiffusive
motion, corresponding to particle caging inside the dense glassy
domains. Whereas the cage escape is strongly subdiffusive in aging 
glasses, here we observe instead a remarkable {\it
superdiffusive} particle motion, with $\Delta^2(t,t_w) \approx
(t-t_w)^{1.6}$, eventually crossing over to a diffusive regime at
very large times. This superdiffusive regime is insensitive to 
the choice of damping in the DPD thermostat, demonstrating that it is a
robust physical feature of the aging process. 

\begin{figure}
\centerline{\includegraphics[width=8cm]{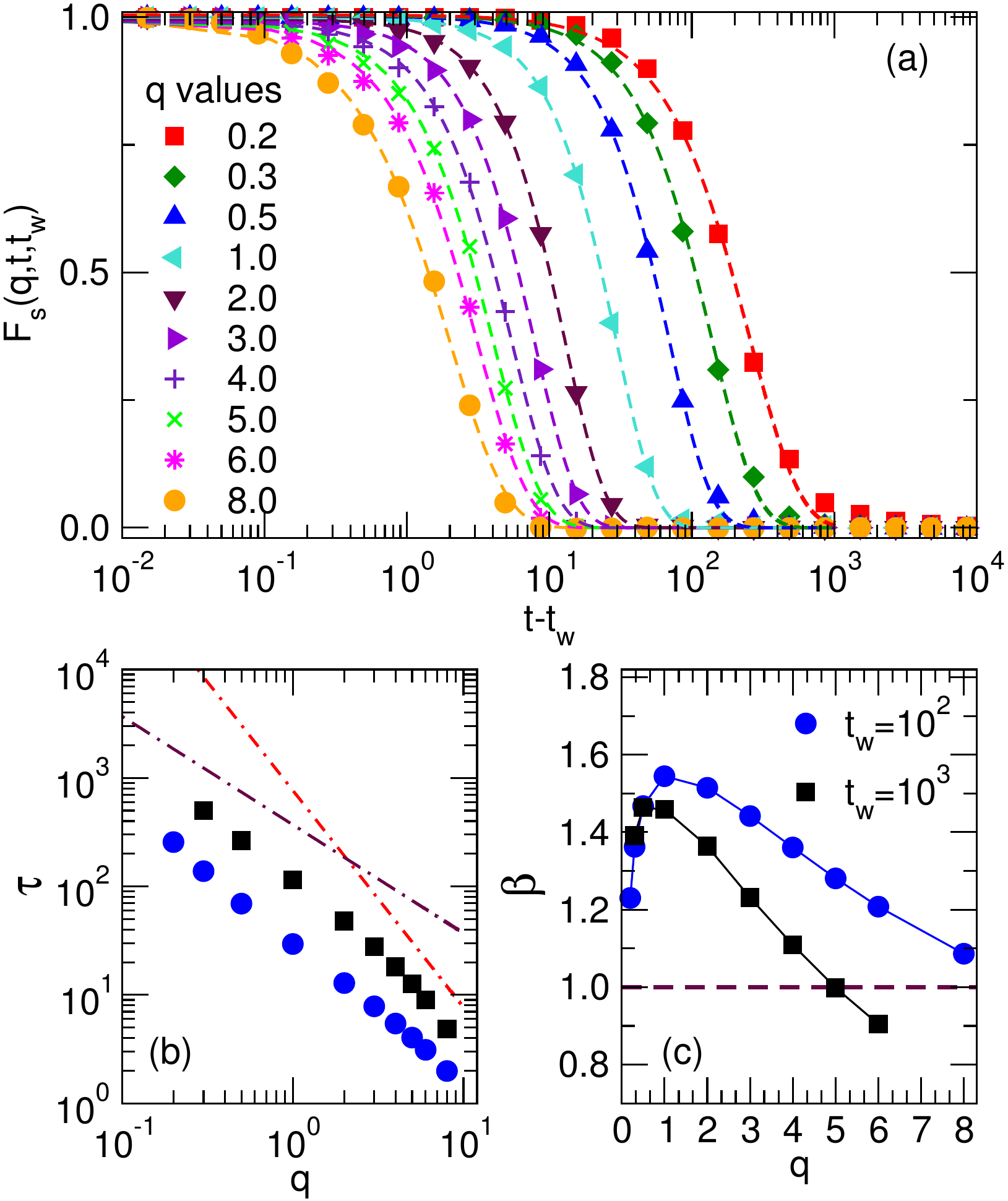}}
\caption{(a) $F_s(q,t,t_w)$, for different
wavevectors $q$, for $t_w=10^2$, $T=0.15$ and $\rho=0.5$. The dashed lines 
correspond to compressed exponential fits with 
fitting parameters reported in (b, c) for both 
$t_w=10^2$ (circles) and $10^3$ (squares).}
\label{fig3}
\end{figure}

To get closer to experimental observations, we spectrally resolve
the dynamics and consider the self-part of the intermediate scattering
function $F_s(q,t,t_w) = \frac{1}{N_A} \sum \left\langle \exp \left\{
i {\bf q} \cdot [ {\bf r}_i(t) - {\bf r}_i(t_w)] \right\} \right\rangle$
for a given scattering vector ${\bf q}$. The data in  Fig.~\ref{fig2}(b)
are for $|{\bf q}| = 1$, thus probing the dynamics over a `mesoscopic'
lengthscale of about 5 particle diameters.  The data again display
a strong dependence on $t_w$, with the relaxation slowing down with
$t_w$.  A striking observation is that the time decay of these
correlations can be fitted with a {\it compressed exponential} form,
$F_s(q,t,t_w) \approx A \exp[- ( \frac{t-t_w}{\tau} )^\beta]$ with
fitting parameters $(A,\beta,\tau)$ which depend both on the age
$t_w$ and on the wavevector $q$.  The fitted $\beta$-values in
Fig.~\ref{fig2}(b) indicate that $\beta > 1$. This implies that
there is a fast relaxation process emerging at mesoscopic length
scales in the aging gel. This corresponds also to the superdiffusive
regime in Fig.~\ref{fig2}(a). We suspect that finite-size effects
affect data for the largest $t_w$ shown.

We also investigate how the relaxation varies across lengthscales,
by studying the variation of  $F_s(q,t,t_w)$ with $q$ in
Fig.~\ref{fig3}(a) for a given age, $t_w=10^2$.  The corresponding
values for the relaxation time $\tau$ and the exponent $\beta$
are shown in Figs.~\ref{fig3}(b,c) for two waiting times. In both
cases, a complex behaviour is observed.  At large $q$, corresponding
to in-cage motion, we find diffusive dynamics $\tau \sim q^{-2}$
accompanied by stretched exponential relaxations, $\beta < 1$.  At
larger lengthscales, a crossover to nearly ballistic dynamics $\tau
\sim q^{-1}$ is accompanied by compressed exponential $\beta >1$
with a peak value near $\beta \approx 1.5$ as frequently reported
in experiments \cite{ranjiniprl2004, duriprl2009}.  Eventually,
diffusive dynamics and exponential relaxation should be recovered
at large enough lengthscales, but these are difficult to access
within our numerical simulations.

\begin{figure}
\centerline{\includegraphics[width=8.8cm]{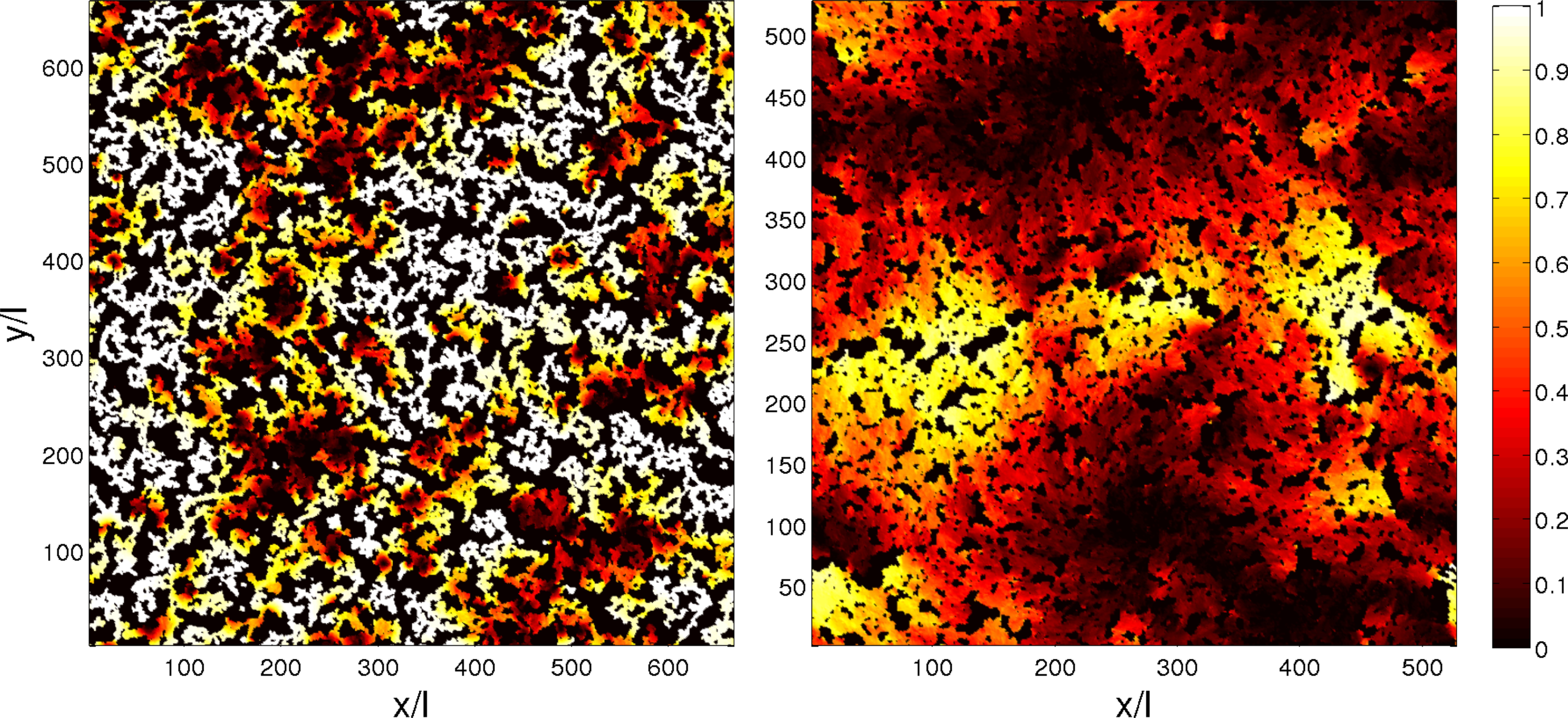}}
\caption{For $\rho=0.5$ (left) and $0.8$(right), spatial map of 
the mobility field, ${q}(x,y)$, at $t-t_w=482.74$, 
for $t_w=10^2$.}
\label{fig4}
\end{figure}

To visualise the spatio-temporal evolution of the relaxation, we
resolve the dynamics for each particle using a mobility function
$q_i(t,t_w)=1-\exp[-{\Delta_i^2(t,t_w)}/2a^2]$, such that $q_i(t,t_w)
\approx 1$ when the particle's squared displacement $\Delta_i^2(t,t_w)
\gg a$, and $q_i(t,t_w) \approx 0$ otherwise. We then construct a
map of the mobility field $q({\bf r},t,t_w)$ by creating a spatial
grid of mesh size 1.5, and averaging $q_i$ over each grid for a
given pair $(t,t_w)$.  Varying the probe length $a$ allows us to
independently consider the various regimes of single-particle
dynamics.  Our most striking finding is reported in Fig.~\ref{fig4},
where we consider the mobility field for $a=6$ corresponding to the
superdiffusive regime and compressed exponential dynamics.  These
maps reveal that, during the aging process, the local relaxation
dynamics is highly heterogeneous and becomes spatially correlated
over {\it ultra-long lengthscales} which are much larger than the
typical domain size characterizing the gel.  We find no such large
dynamic correlation length when the probe length $a$ is smaller,
since then only the caged glassy dynamics is probed. Such features
are again in excellent agreement with experiments.

\begin{figure}
\centerline{\includegraphics[width=8.5cm]{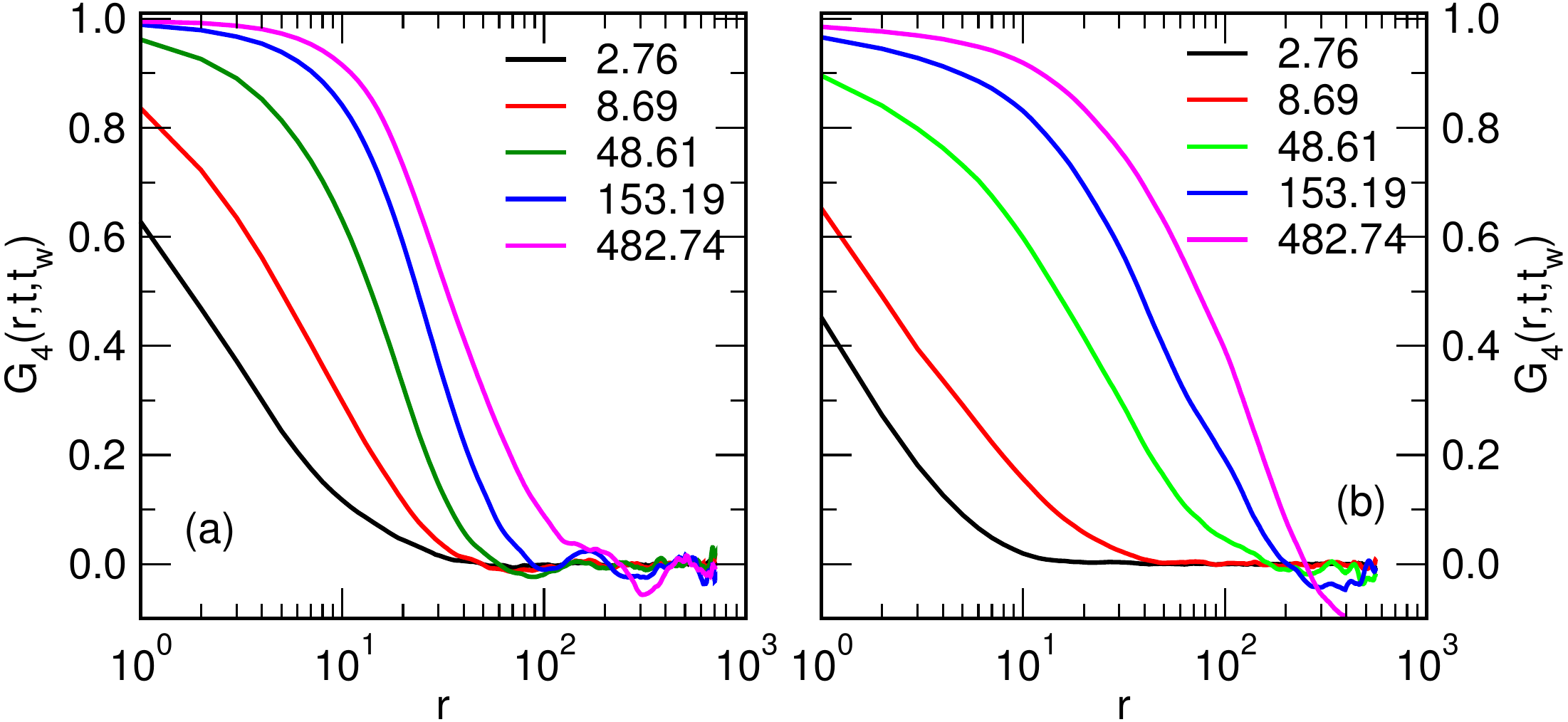}}
\centerline{\includegraphics[width=8.5cm]{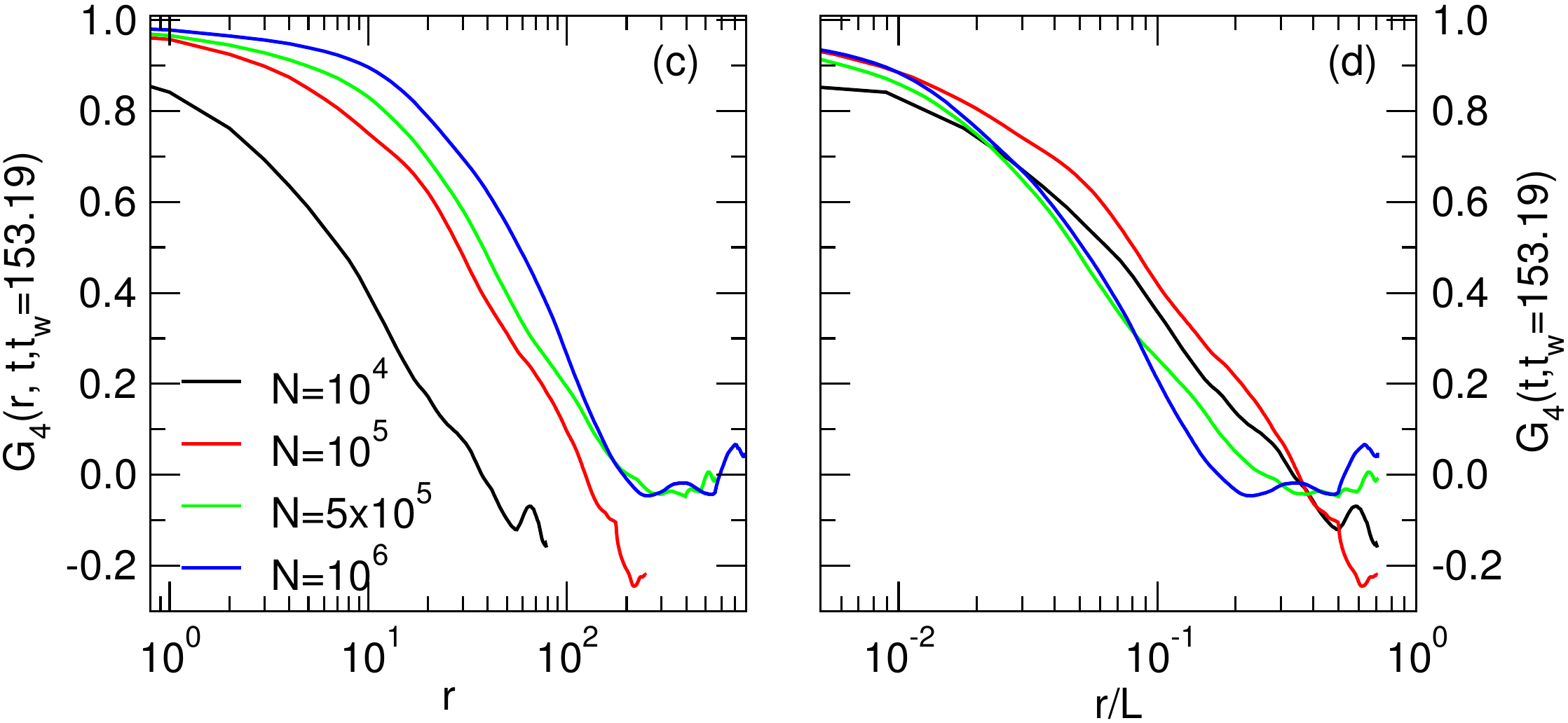}}
\caption{ Spatial correlations of mobility, $G_4(r,t,t_w)$, 
for different $t-t_w$ (as marked), corresponding to $t_w=10^2$,  at 
$\rho=0.5$ (a), $0.8$ (b), and (c) for different system sizes ($N$), 
as marked, at $\rho=0.8, t_w=153.19$. 
(d) Variation of $G_4(r,t,t_w)$ corresponding to (c) with rescaled 
distances $r/L$. }
\label{fig5}
\end{figure}

To quantify the size of these dynamic domains, we measure the spatial
correlation function $G_4(r,t,t_w)=\langle\delta{q}({\bf
r'},t,t_w)\delta{q}({\bf r+r'},t,t_w) \rangle/ \langle\delta{q}^2({\bf
r'},t,t_w)\rangle$, where $\delta q = q - \overline{q}$, with
$\overline{\cdots}$ corresponding to a spatial average.  Representative
data for $G_4(r,t,t_w)$ are shown in Fig.~\ref{fig5}. We first show
in Figs.~\ref{fig5}(a,b) that the size of correlated domains increases
with the time $t-t_w$ indicating that large-scale correlations
develop as the structure reorganises.  In agreement with Fig.~\ref{fig4},
the correlations seem longer for the larger density. Such large
length scales (of more than 100 particle diameters)
measured in our simulations are reminiscent of the
ultra-long correlation lengths measured in experiments using spatially
resolved light scattering techniques~\cite{duriepl2006}.

In addition, we also observe that the measured correlation lengths,
for any $t-t_w$, depend upon the system size. This is seen in
Fig.~\ref{fig5}(c), where we plot $G_4(r,t,t_w)$ for various system
sizes at a given state point. Clearly, the correlation functions
decay more slowly for larger systems. If we rescale distances by
the linear size of the simulation box, $L$, there is an approximate
collapse of $G_4(r,t,t_w)$; see Fig.~\ref{fig5}(d). Such a dependence
of the correlations with $L$ suggests that the system size is the
only characteristic lengthscale over which the relaxation is spatially
correlated, at least for the system sizes that we can access
numerically (our largest $L$ is about 2000 particle diameters).
Direct visualization indicates that such large scale correlations
arise due to the spontaneous elastic recoiling of the network over
large distances after local bonds intermittently break due to thermal
activation.

Overall, our large-scale simulations reproduce all anomalous features
observed in scattering experiments performed in aging gels, most
notably subdiffusive caged dynamics crossing over to superdiffusive,
compressed exponential dynamics at large enough lengthscales
characterized by ultra-long-ranged dynamic correlations. In our
model, this arises spontaneously 
because rare localised relaxation events
affect the dynamics of the tenuous gel structure over
large lengthscales. Such aging dynamics is not observed in dense
glasses.  Although our observations are for a two-dimensional
system, the underlying mechanism would also apply for three-dimensional
experimental systems, and future numerical work would explore that.
Studying the mechanical response of such aging nonequilibrium gels
would also be interesting, with a possible interplay of intrinsic and
extrinsic timescales. Also, the potential connection to the observation
of sudden collapse in many gels after an initial lag time
\cite{bartlettpre2012} needs exploration. Whether the
mechanisms discussed in the context of such low density materials
can be extended to understand similar fast relaxations reported 
for denser aging glasses \cite{bob1,ruta1,ruta2} also requires
more investigations.

\acknowledgments 
We thank L. Cipelletti for useful discussions and the HPC  at IMSc
for providing the computational facilities.  The research leading
to these results has received funding from the European Research
Council under the European Union’s Seventh Framework Programme
(FP7/2007–2013)/ERC Grant Agreement No. 306845.

\end{document}